\documentclass[aps,physrev,preprint,groupedaddress]{revtex4-2}
\usepackage[english]{babel}
\usepackage{graphicx}
\usepackage{epstopdf}

\begin{document}


\title{\textbf{Impact of metallographic polishing on the RF properties of Niobium for SRF applications.} 
}%

\author{Oleksandr Hryhorenko}
 \email{Contact authors: hryhoren@jlab.org, david.longuevergne@ijclab.in2p3.fr, oliver.kugeler@helmholtz-berlin.de, sebastian.keckert@helmholtz-berlin.de}
\affiliation{%
 Thomas Jefferson National Accelerator Facility (JLAB), 12000 Jefferson Ave, 23606 Newport News, VA, United States
}%
\author{Anne-Marie Valente-Feliciano}%
\affiliation{%
 Thomas Jefferson National Accelerator Facility (JLAB), 12000 Jefferson Ave, 23606 Newport News, VA, United States
}%

\author{David Longuevergne}
\affiliation{%
Université Paris-Saclay, CNRS/IN2P3, Laboratoire de Physique des 2 Infinis Irène Joliot-Curie (IJCLAB), 15 rue Georges Clémenceau, 91405 Orsay, France}%

\author{Claire Zylberajch Antoine, Thomas Proslier, and Fabien Eozenou}
\affiliation{%
Université Paris-Saclay, CEA Département des Accélérateurs\\ de la Cryogénie et du Magnétisme (CEA-IRFU), 91191 Gif-sur-Yvette, France}%

\author{Oliver Kugeler}
\affiliation{%
Helmholtz-Zentrum Berlin f\"ur Materialien und Energie GmbH, Hahn-Meitner-Platz 1, 14109 Berlin, Germany
}%

\author{Sebastian Keckert}
\affiliation{%
Helmholtz-Zentrum Berlin f\"ur Materialien und Energie GmbH, Hahn-Meitner-Platz 1, 14109 Berlin, Germany
}%

\author{Jens Knobloch}
\affiliation{%
Helmholtz-Zentrum Berlin f\"ur Materialien und Energie GmbH, Hahn-Meitner-Platz 1, 14109 Berlin, Germany
}%
\affiliation{%
Universität Siegen, Department Physik, Walter-Flex-Str.\ 3, 57068 Siegen, Germany
}%

\date{\today}

\begin{abstract}

The performance of superconducting radio-frequency (SRF) cavities made of Niobium is tied to the quality of their inner surfaces exposed to the radio frequency (RF) waves. Future superconducting particle accelerators, because of their dimensions or the unprecedented stringent technical requirements, require the development of innovative surface processing techniques to improve processing reliability and if possible ecological footprint and cost, compared to conventional chemical processes. Metallographic polishing (MP) has emerged as a promising polishing technology to address these challenges. Previous studies focused on the characterization of the processed material surface at room temperature in the absence of RF waves. However, the evaluation of material properties, such as surface resistance under RF, at cryogenic temperature has failed, primarily due to the unavailability of devices capable of achieving the necessary resolution in the nanohm range. To overcome this limitation, a quadrupole resonator (QPR) has been utilized. The RF results demonstrate that the MP polishing, developed to preserve a high-quality niobium surface with very low surface resistance, is highly effective compared to conventional polishing. This conclusion is further supported by topography and microstructural analysis of the QPR top-hat samples, which revealed the clear superiority of the metallographic approach.

\end{abstract}


\maketitle


\section{\label{sec:intro}Introduction}
The superconducting radio frequency (SRF) cavity is the core component in high-power particle accelerators ($\sim$MW)  that deliver high-energy ($\sim$TeV) and high-current beams ($\sim$mA) \cite{Padamsee_2017, ess2024, cebaf24, singer16, Dhakal24, GONNELLA2018143, CALATRONI200695}. These cavities are typically made of bulk niobium (Nb) or copper coated with a niobium film \cite{ess2024, Singer_2017, CALATRONI200695, Sladen2010STATUSOS}. The fabrication process of these cavities is complex and expensive, and results in the formation of an imperfect inner layer known as the "damage layer" \cite{Antoine:2012hna}. This layer causes significant RF dissipation and should be removed to reach high-quality ($Q$) and high-gradient operation ($E_{acc}$).
Methods such as Buffered Chemical Polishing (BCP) and Electropolishing (EP) are routinely used to etch the inner surface of the cavities of about 150 µm. These techniques aim to create a smooth, chemically clean, crystal-damage-free surface to sustain intense RF waves. Both BCP and EP use hazardous, HF-based acids and involve etching or diffusion-limited polishing processes, resulting in varied surface finish \cite{TIAN20061236, hui2010}. Moreover, surface defects, such as scratches or pits, tend to slowly worsen during the chemical processing,  contributing to reliability issues. Local grinding and Centrifugal Barrel Polishing (CBP) are utilized to address significant defects \cite{iwashita:srf19-thp088, Cooper_2013}. However, even after these processes, the surface still requires BCP or the light EP to eliminate any remaining defects and abrasive contaminants. 

Due to the large cavity numbers for future large-scale accelerators like the International Linear Collider (ILC) and the Future Circular Collider (FCC) \cite{Aryshev2021, fcc2023}, high and reliable production yield, fabrication cost reduction, and increased cavity performance are required.
Hence, alternative surface preparation and cavity fabrication techniques are being investigated to tackle the challenges associated with these demands. Moreover, alternative superconductors (Nb$_3$Sn, MgB$_2$, etc...) are also under study as stand-alone layers and nanometric superconductor-insulator-superconductor (SIS) structures in the community to go beyond bulk Niobium limits \cite{Posen_2017, Keckert_2019, He_2012, kubo_2014}. These thin film structures require defect-free and smooth substrates as the film quality is highly dependent on the substrate quality \cite{kubo_2014}. To address the challenges associated with final cavity RF performance and/or substrate quality, we have developed metallographic polishing (MP) methods as reported in Ref. \cite{jmmp_23}.

Previously, surface characterization was limited to laser confocal microscopy, scanning electron microscopy, EBSD, EDS, and SIMS spectroscopy to examine the material properties at room temperature \cite{jmmp_23, Hryhorenko:2019own, Hryhorenko:2023hin}. One first attempt of cryogenic characterization under RF using a "mushroom" cavity type was performed in 2019 \cite{hryhorenko:srf19-thp002, Hryhorenko:2019own}, but the results turned out to be non-exploitable. Indeed, the sample could not be heat treated to degas hydrogen, resulting in a significant degradation of the surface resistance due to Niobium hydride precipitation (so called Q-disease) \cite{Knobloch_2003}, and high-frequency operation \cite{welander:srf15-tupb065, guo:pac11-tup102} leading to poor resolution in the micro-ohm range \cite{hryhorenko:srf19-thp002}. 


In this paper, we will describe an application of the revised sequence for the top-hat quadrupole resonator (QPR) sample and will present measurements performed with this state-of-the-art device to fully qualify the MP-processed Niobium surface under RF and at cryogenic temperatures, with nano-Ohm resolution \cite{qpr_hzb_2021_rsi, qpr_hzb_2021_aip}. We also characterize surface alterations resulting from conventional chemical polishing and the MP polishing process, with a focus on topography and microstructural changes. This study highlights the significance of surface feature size and crystal quality in influencing RF performance.

\section{\label{sec:device}RF measurement method and sample preparation}
 
\subsection{\label{sec:device:QPRtech} QPR cavity presentation and methodology}
The Quadrupole Resonator (QPR) is a dedicated sample test cavity that enables direct measurements of a sample's surface resistance at RF frequencies in a wide parameter space of sample temperature and RF field level \cite{qpr_hzb_2021_rsi}. The resonator features three quadrupole modes with frequencies near 415, 847 and 1290 MHz that provide an RF magnetic field of up to 120 mT on the sample surface \cite{kleindienst:srf15-wea1a04}. While the resonator is kept at constant temperature, stabilized by a superfluid liquid helium bath at 1.8 K, the sample is thermally decoupled from the resonator and can be heated to arbitrary temperatures. Using a PID control loop for the heater power and a calibrated temperature sensor on the sample, any RF dissipation on the sample is observed as heater power difference yielding a calorimetric measurement of the surface resistance ($R_{\text{S}}$) at a given RF field level. Note that this measurement is independent of any losses occurring in the cavity, hence there is no need to calibrate the surface resistance measurement to a reference sample.
Measuring temperature-dependent $R_{\text{S}}$ gives access to the residual resistance ($R_{\text{res}}$) using the common formula $R_{\text{S}}(T)=R_{\text{BCS}}(T)+R_{\text{res}}$, where $R_{\text{BCS}}$ is the BCS surface resistance.
For a more detailed description of the QPR system the reader is referred to Ref. \cite{qpr_hzb_2021_rsi}.

\subsection{\label{sec:device:QPRsample}QPR sample description and baseline sample preparation}

The dimensions of the QPR sample make it compatible with various existing coating facilities, making these samples valuable for studying how deposition techniques and their parameters affect the RF performance of the deposited films. Moreover, the QPR sample can also be utilized for studies of surface processing techniques, as demonstrated in subsequent sections of this work. 

The QPR sample consists of a flat niobium disk (RRR=300) with a diameter of 75 mm and an initial thickness of 10 mm. This disk is electron beam welded to a niobium cylinder, which has a height of 95 mm. The cylinder has a top-hat like rim of 99 mm diameter at its bottom with which the sample is mounted to a stainless steel conflat flange (designated as CF100) to facilitate attachment to the host QPR cavity. The flange is niobium-coated to minimize parasitic losses, hence enabling the measurement of surface resistance down to the sub-nano-Ohm range \cite{qpr_hzb_2021_aip}. This design allows for easy mounting and dismounting of the sample from the host cavity, with only the flat niobium disk being exposed to high radio frequency (RF) fields.

The QPR sample was polished using the standard BCP solution, which consists of hydrofluoric acid (HF-49\%), nitric acid (HNO$_3$-70\%), and phosphoric acid (H$_3$PO$_4$-85\%) in a volume ratio of 1:1:2. Subsequently, the QPR sample underwent bulk EP to remove 100 µm of material, using a mixture of HF (40\%) and sulfuric acid (H$_2$SO$_4$ at 97\%) in a 1:9 volume ratio. The EP was performed in a static electrolyte at an applied voltage of 6 V. The electrolyte temperature was gradually increasing from an initial 5 °C to a maximum of 30 °C by the end of the process. The sample was then annealed under vacuum at 900\textdegree C for 3 hours followed by a light EP. During the final EP, the QPR was electropolished at 6V and below 25 °C. All these standard processes were performed at CEA-Saclay, on the platform Le Synergium \cite{synergium2025}. An RF test was then performed at HZB with the QPR facility to establish the baseline performance of the material. We refer to these results as the baseline measurement in the following section. 

\subsection{\label{sec:device:polishing}Sample Preparation by MP}

Subsequently, the same QPR sample was MP-polished using a LAM PLAN Masterlam 1.0 polishing device with LAM PLAN consumables in operation on Vide\&Surface platform at IJCLab \cite{vide_et_surface2025}. This process involved a two-step procedure documented in \cite{jmmp_23}. 
The first step (lapping step) aimed to planarize the surface and improve roughness, removed 20 µm of material. A rigid composite disk with 3 µm polycrystalline diamonds was used. 
The second step (polishing step) focused on recovering purity and microstructure by removing the layer damaged and contaminated by the previous step. Although this step increases roughness due to re-appearance of grains, it is essential for the recovery of optimal superconducting properties. In the second step, a polyurethane cloth combined with a solution based on the silica colloidal SiO$_2$ (50 nm), peroxide H$_2$O$_2$, and ammonia (NH$_4$OH), diluted in deionized water (up to 20\%) was used to remove approximately 5 µm.
After polishing, the sample was annealed at 600°C for 10 hours with the vacuum furnace in operation on Supratech platform at IJCLab \cite{supratech2025} to remove hydrogen contamination. A high-pressure rinse (HPR) was performed to remove colloidal silica or other contaminants from the furnace. Finally, the sample was remeasured under RF at HZB.

\section{\label{sec:results}Results and discussion}

\subsection{\label{sec:results:topography}Sample topography and microstructure analysis}

\begin{figure}
    \centering
    \includegraphics[width=1\linewidth]{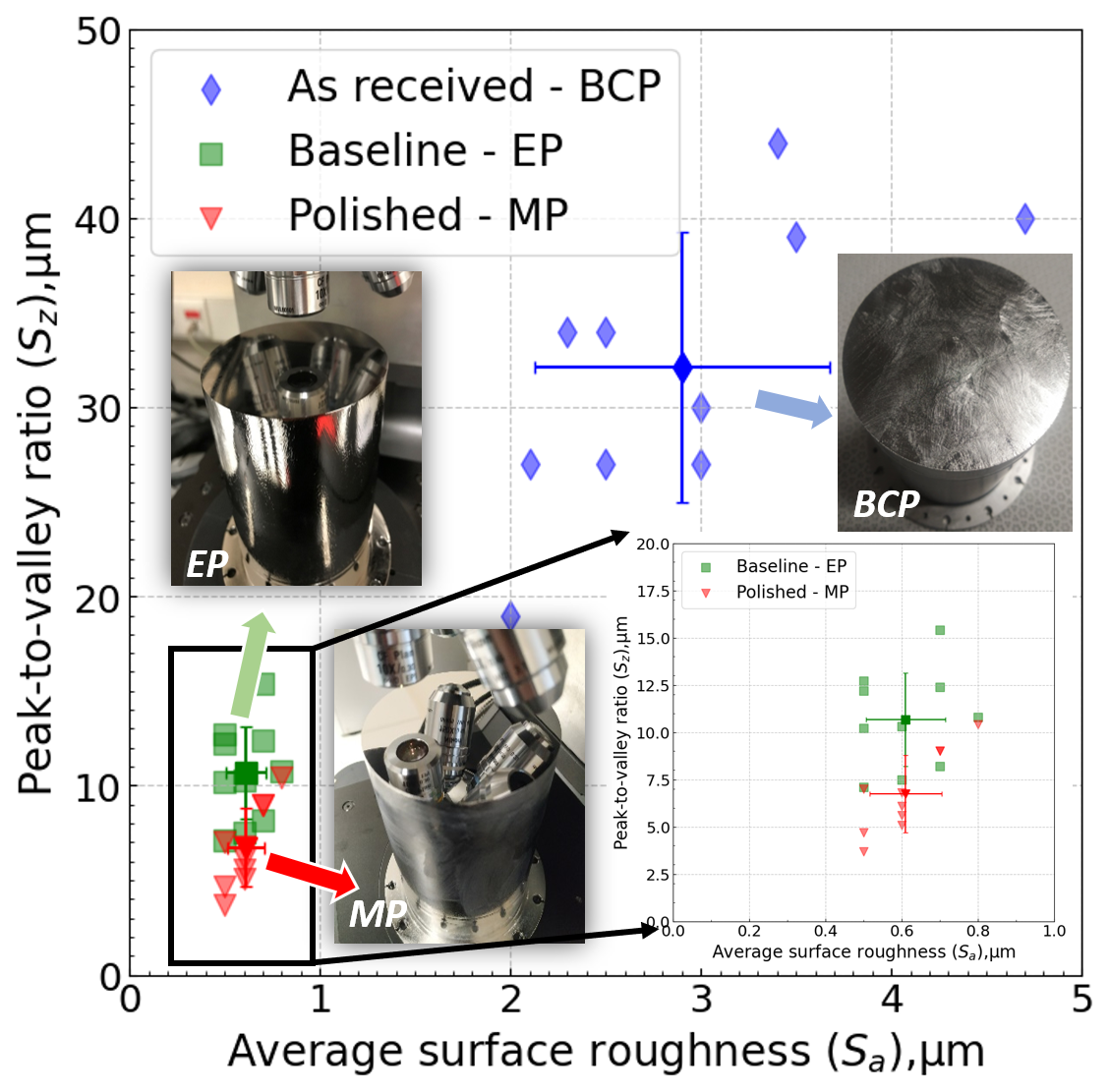}
    \caption{Peak-to-valley ratio and average surface roughness comparison depending on the surface processing. Note that the scales of the axes are not equal. The BCP surface exhibits low reflectivity, while the MP surface is highly reflective.}
    \label{fig1}
\end{figure}

The QPR sample surface quality was examined at each treatment step using laser confocal microscope (Keyence VKX 210) of Vide\&Surface platform to study how the different surface processing methods -- BCP, EP, and MP -- affect the topography of the QPR surface. The peak-to-valley roughness ($S_z$) and average roughness ($S_a$) vary with each processing method, as illustrated in Figure \ref{fig1}. Measurements indicate significant variations in topography after BCP surface processing (as received) due to the etching mechanism, which reveals surface quality to the underlying surface. The peak-to-valley roughness $S_z$, and average surface roughness $S_a$ after BCP are 32.1$\pm$7.2 µm and 2.9$\pm$0.8 µm, respectively. The surface tailoring with the baseline EP processing considerably improved the topography, $S_z$=10.7$\pm$2.5 µm, and $S_a$=0.6$\pm$0.1 µm. Additionally, further improvements in topography are observed after MP processing, $S_z$=6.7$\pm$2.0 µm, and $S_a$=0.6$\pm$0.1 µm. 

\begin{figure}
    \centering
    \includegraphics[width=1\linewidth]{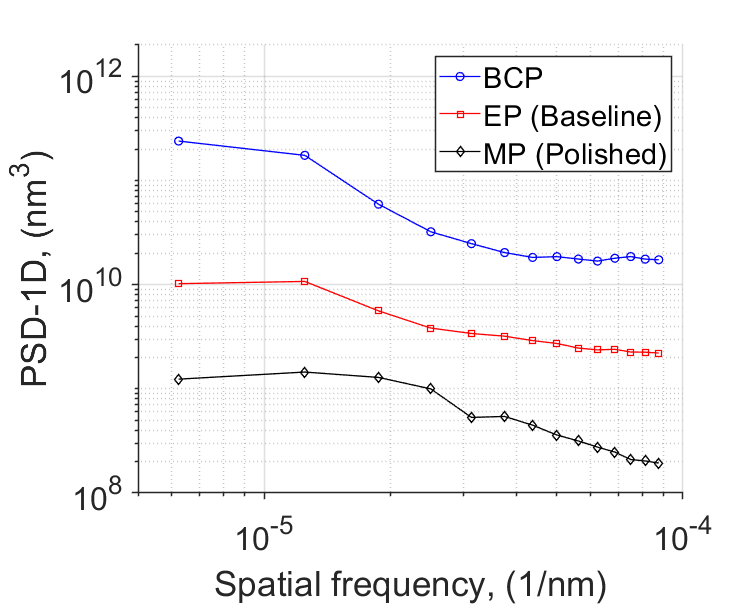}
    \caption{Average power spectral density spectra for various surface processing methods (BCP, EP, and MP) of the QPR sample, derived from laser confocal data. The graph illustrates the contribution of different surface features across multiple spatial frequency domains. }
    \label{fig2}
\end{figure}

In addition to the reported roughness values \( S_a \) and \( S_z \), we calculated the power spectral density (PSD) spectra by performing a Fourier transform on the height data obtained from laser confocal scans \cite{xu2011}. We conducted a total of ten scans over an area of $1000\times1000$\,µm. The values of \( S_a \), \( S_z \), and the PSD are influenced by both the scan resolution and the size of the scanned area. While the PSD provides insights into the distribution of the height signal across various spatial frequencies of the surface, the \( S_a \) and \( S_z \) values do not offer this type of information. Figure \ref{fig2} shows the distribution of surface features intensities across the spatial frequency range, indicating that the MP-polished is superior across all frequencies. 

\begin{figure}
    \centering
    \includegraphics[width=1\linewidth]{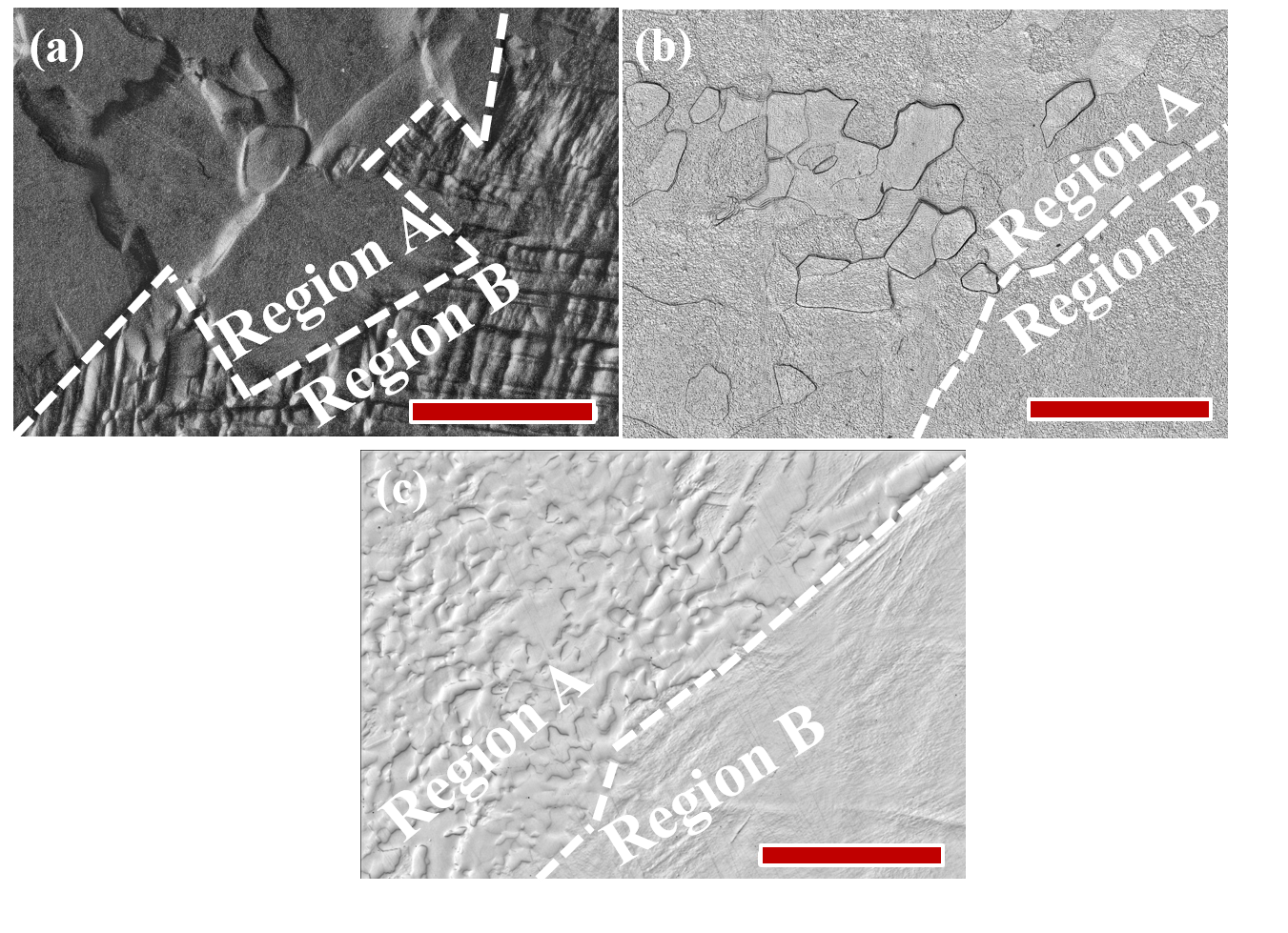}
    \caption{Typical laser confocal images of the bulk QPR after different surface processing: (a) BCP, (b) EP, and (c) MP. Scale bars are 1000 µm. Region A is separated from Region B with a dashed line.}
    \label{fig3}
\end{figure}

All polishing techniques result in a considerable variation in the surface finish, with some areas being significantly smoother than others. Figure \ref{fig3}a illustrates that the niobium grain structure is not uniform after BCP, as previously reported \cite{tikhonov:srf19-tup073}, but the cause for this non-uniformity was not mentioned. As shown in Fig. \ref{fig2}b,c similar results are observed after EP and MP surface treatments. Even after removing approximately 500 µm of material, some areas exhibit patchy regions with grain formation (Region A), resulting in increased roughness. In contrast, some areas show limited to no grain appearance, resulting in lower roughness (Region B). The extent of cold work applied to the Nb disk before polishing is still unclear, and we speculate that it may be the root cause of this difference. The initial thickness of 10 mm, while typical niobium sheets used for SRF are only 3 mm, could contribute to the non-uniform microstructure due to the varying amounts of cold work, particularly the absence of the rolling step. This variation may lead to an irregular distribution of internal residual stress during the bulk Nb preparation \cite{bieler:srf09-tuoaau03, grill:srf11-thpo059}. High residual stress causes grains in Region B to show slight recovery from cold work without visible recrystallization. In contrast, recrystallization occurs in areas with lower levels of residual stress, as seen in Region A.  

\begin{figure}
    \centering
    \includegraphics[width=1\linewidth]{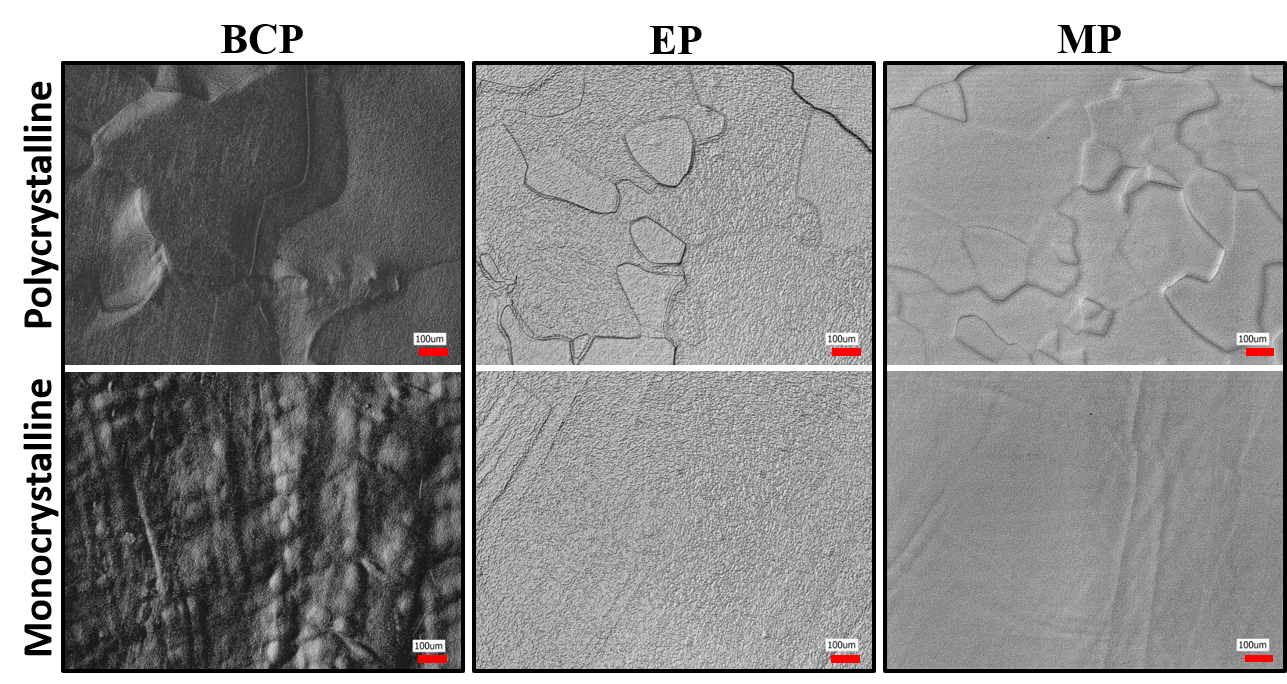}
    \caption{Digital interference contrast (DIC) images illustrating the impact of different polishing procedures on polycrystalline (Region A) and monocrystalline (Region B) Nb grains. The scale bar represents 100 µm, highlighting the differences in grain structures between the two regions as discussed in the text.}
    \label{fig4}
\end{figure}

Region A can be classified as polycrystalline, consisting of many grains with varying orientations that exhibit different surface modifications during processing, as shown in Fig. \ref{fig4}. Region B is predominantly monocrystalline (single grain), resulting in a more uniform response to identical processing methods, although it clearly displays the presence of residual strains. BCP is particularly sensitive to grain orientations due to its etching mechanism, which is affected by variations in surface energy related to crystalline orientation, caused by dislocations \cite{gutman1998mechanochemistry}. The EP process operates through a diffusion-limited mechanism involving the diffusion of fluorine ions through a viscous layer \cite{hui2010}. This results in a smoother surface compared to BCP, however, some subsurface damage is present within the grain structure, primarily due to the presence of residual strains. The MP process further enhances surface smoothness while minimizing damage, primarily through processes like passivation and oxide growth. Overall, the effects of the baseline surface processing method (EP) and MP are examined through measurements of surface resistance at different surface magnetic fields, as discussed in the next chapter.

\subsection{\label{sec:results:Rs}Surface resistance analysis}


The QPR measurements were performed for the baseline sample (100 µm EP, annealing at 900$^{\circ}$C for 3 hours, and light EP 20 µm) and for the MP polished sample at frequencies of 415 MHz and 847 MHz.

\begin{figure}
    \centering
    \includegraphics[width=1\linewidth]{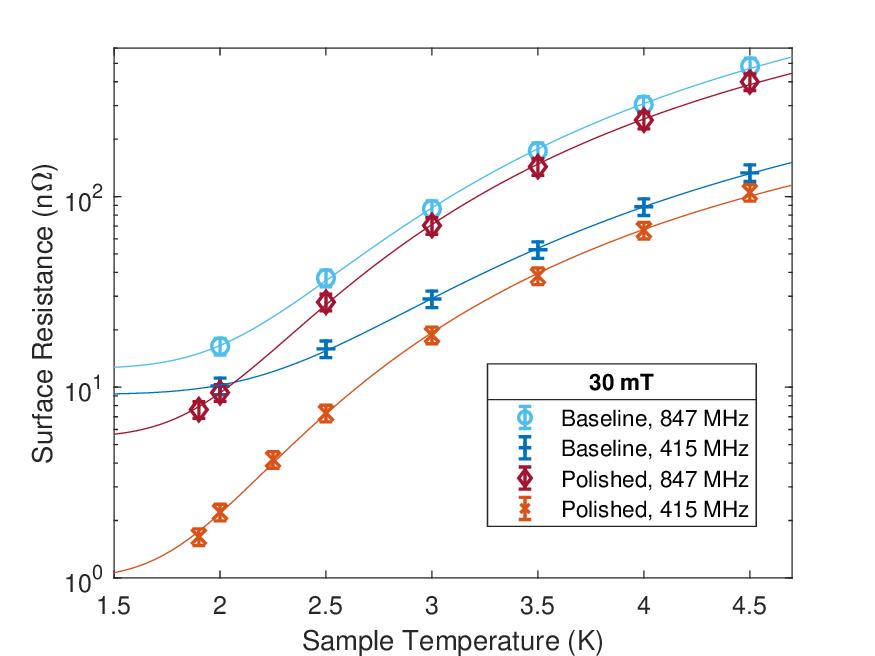}
    \caption{Surface resistance measurement versus temperature for baseline and MP sample at different frequencies for an RF magnetic field of 30 mT. Solid lines show the fits and the extrapolation to 0\,K that is used to extract the residual resistance.}
    \label{fig5}
\end{figure}

Fig. \ref{fig5} shows a typical measurement data set of surface resistance vs. sample temperature for baseline and MP sample at two different frequencies and for an RF magnetic field of 30 mT. From these curves, the residual resistance is extracted using the phenomenological approximation $R_{\text{S}}=\frac{a}{T}\exp\left(-\frac{b}{T}\right)+R_{\text{res}}$ for $T$$\le$4.5 K \cite{martinello:srf17-tuyaa02}.

\begin{figure}
    \centering
    \includegraphics[width=0.49\linewidth]{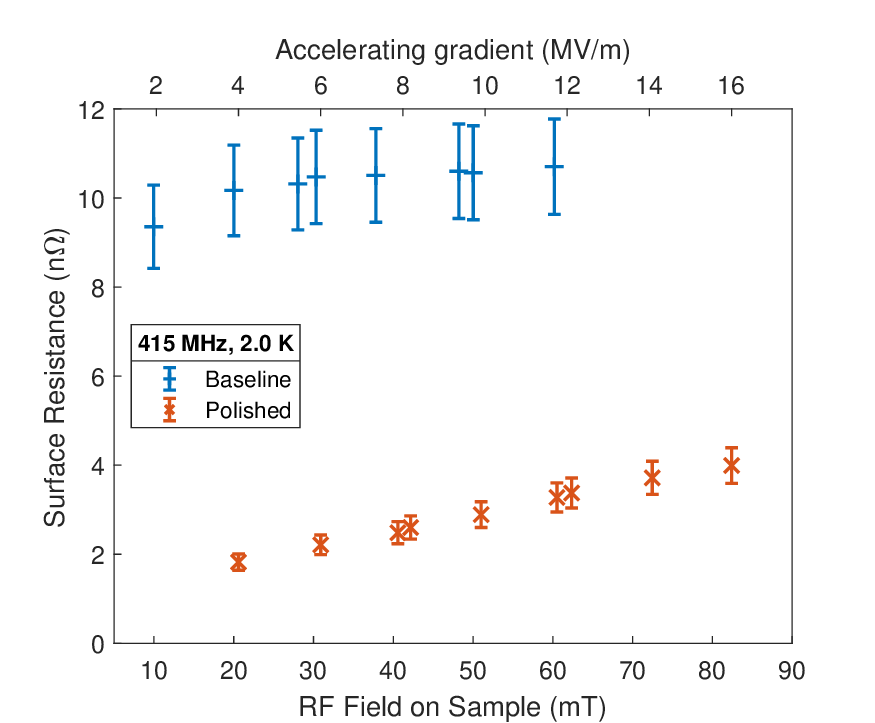}
    \includegraphics[width=0.49\linewidth]{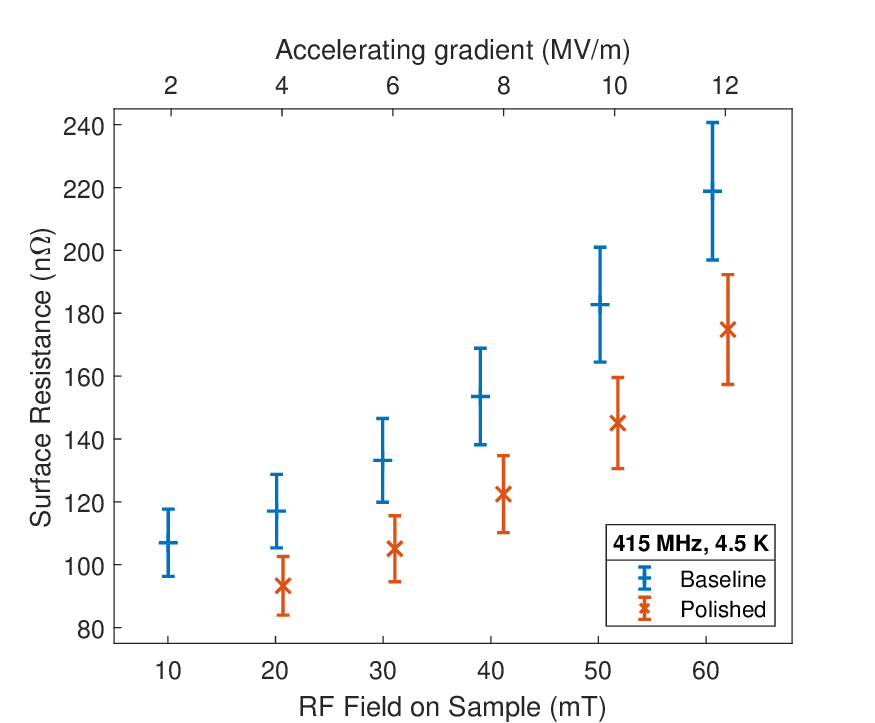}
    \includegraphics[width=0.49\linewidth]{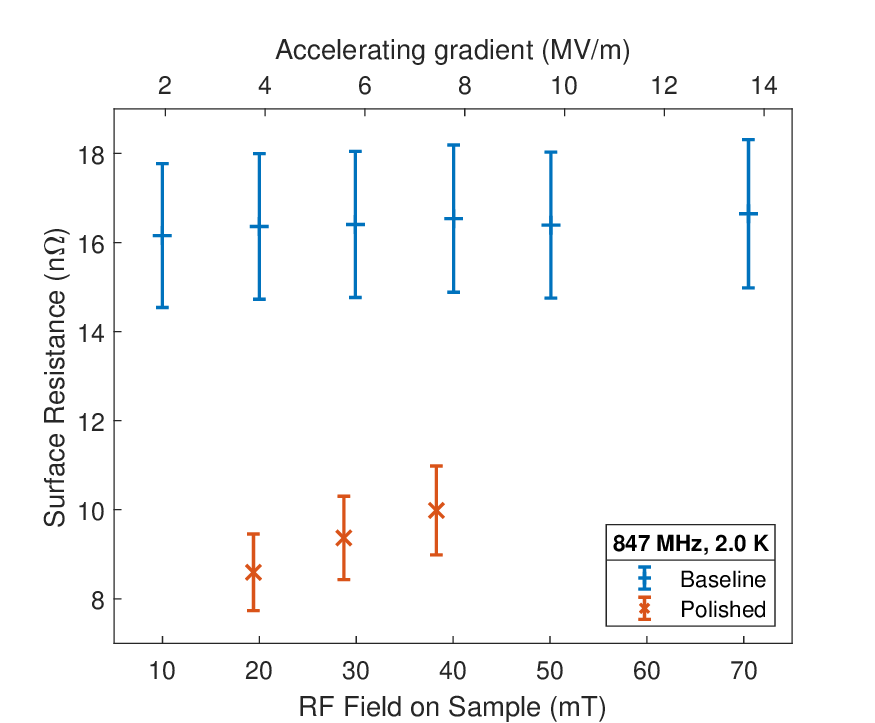}
    \includegraphics[width=0.49\linewidth]{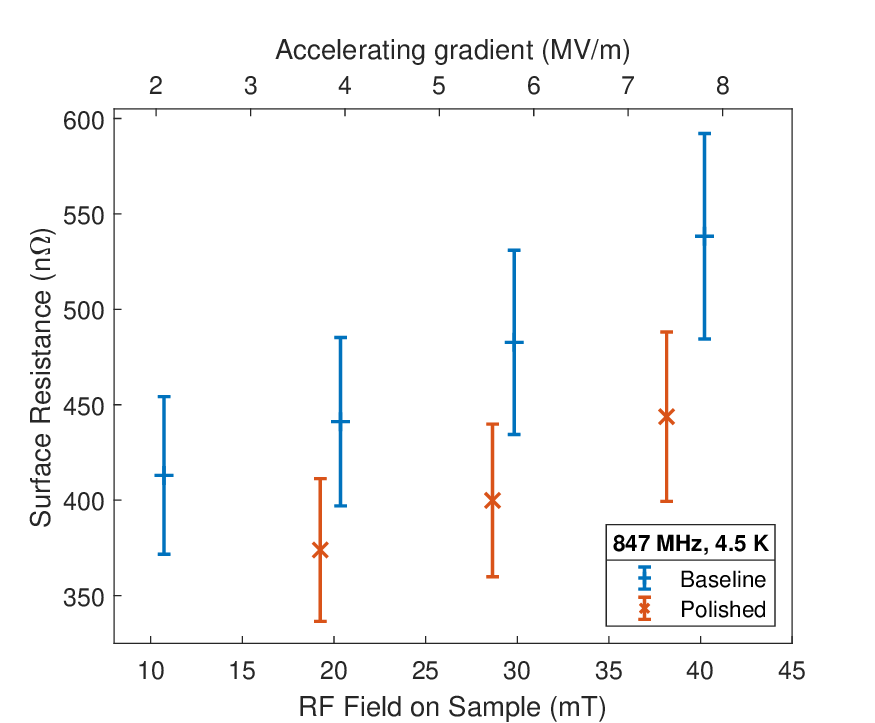}
    \caption{Surface resistance data for baseline and MP sample versus RF magnetic field at different frequencies and sample temperatures. Second x-scales are added showing the equivalent accelerating gradient for an LHC-type cavity.}
    \label{fig6}
\end{figure}

Fig. \ref{fig6} shows various data sets of surface resistance vs. RF magnetic field at two frequencies and at sample temperatures of 2.0 K and 4.5 K.
The second abscissa provides the equivalent accelerating gradient of an LHC-type cavity operating at 400 MHz \cite{Boussard:410377}.
The MP sample shows a significantly lower $R_{\text{S}}$ as compared to the baseline measurement, with a difference between the two that is even larger at higher temperatures.
Hence, both $R_{\text{res}}$ and $R_{\text{BCS}}$ are improved by MP.
The maximum possible RF field level for a given sample temperature typically depends on the RF heating of the sample and hence on the surface resistance. Thanks to the reduced $R_{\text{res}}$, at 415 MHz and 2.0 K a higher field level of more than 80 mT was achievable without a quench being observed.

\begin{figure}
    \centering
    \includegraphics[width=1\linewidth]{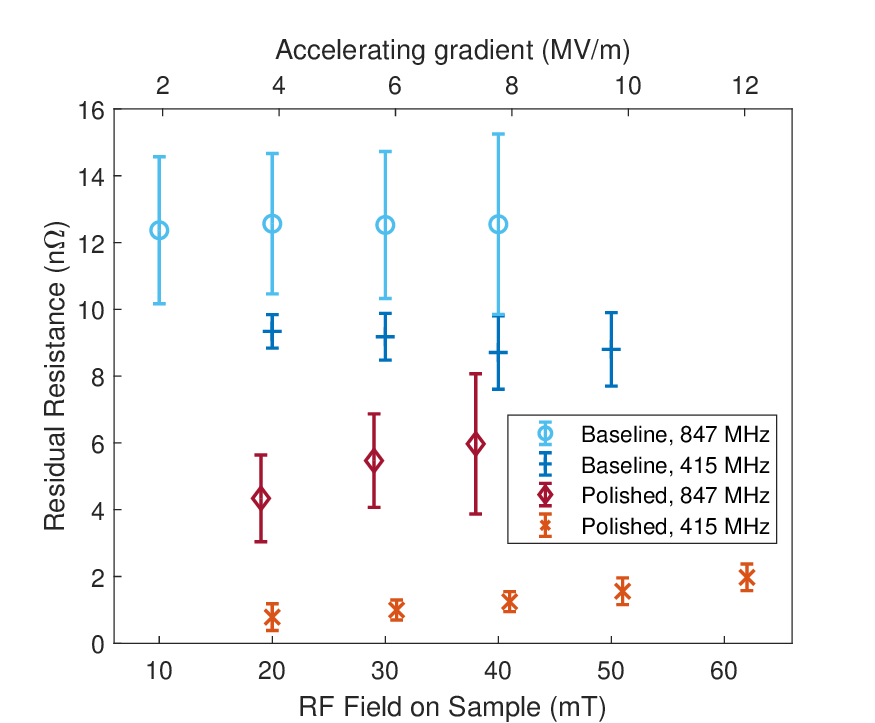}
    \caption{Residual resistance vs RF magnetic field for both samples and RF frequencies. This data is extracted from fitting $R_{\text{S}}(T)$ data sets obtained at fixed field level like those shown in Fig. \ref{fig3}.}
    \label{fig7}
\end{figure}

Figure \ref{fig7} shows the extracted residual resistance vs. RF magnetic field for both samples and RF frequencies. A minimum $R_{\text{res}}$ of 0.8 n$\Omega$ was obtained at 415 MHz for the MP sample. After polishing, a slightly increasing slope of $R_{\text{res}}$ with RF field is observed while the baseline sample showed the opposite effect of constant or decreasing $R_{\text{res}}$ with increasing RF field.

\section{\label{sec:conclusions}Conclusions}
We investigated the impact of different surface treatments on bulk niobium topography by comparing buffered chemical polishing (BCP), electropolishing (EP, our baseline), and a novel “metallographic polishing” (MP) protocol developed specifically for SRF niobium. MP adapts standard metallographic polishing techniques -- used to prepare damage‐free sample surfaces -- so as not only to maximize smoothness but also to eliminate dislocations and other near‐surface crystalline defects. Hryhorenko et al. (2023) demonstrated that MP can treat large niobium areas in fewer steps and with a much faster turnaround than conventional methods \cite{jmmp_23}. We then evaluated surface resistance measurements on QPR samples -- both baseline and MP-treated -- across a range of temperatures, magnetic fields, and RF frequencies. MP reduced the peak-to-valley roughness by 37 \% relative to the EP baseline, a smoother finish that coincided with a lowered residual resistance by 6.5 to 8.5\,n$\Omega$ at both investigated frequencies without any additional chemical treatment. These performance gains cannot be attributed to surface smoothness alone. For instance, classical centrifugal barrel polishing yields very low roughness but leaves a subsurface damage layer, necessitating at least a 20 µm EP removal step to achieve good RF performance \cite{cooper:srf11-weioa02, higuchi:2001}. Acid-free extended mechanical polishing (XMP), another metallographic variant, also reaches high gradients without etching, yet it requires a prohibitively long process time \cite{cooper:srf13-tup060, Cooper_2013}. Our findings underscore that both topographical smoothness and crystalline integrity critically influence achievable field levels. Moreover, MP produces reproducible niobium surfaces, providing strong baseline substrates for subsequent single or multilayer coatings. It also enhances cost-effectiveness by allowing the reuse of expensive QPR samples, as the surface can be easily restored to its high quality after RF evaluation. 

Looking ahead, the transition from MP-processed QPR samples to MP-assisted cavity production presents exciting prospects to enhance cavity performance, improve production reliability, minimize ecological footprint, and prioritize worker safety. By prioritizing sheet polishing before cavity forming, we can mitigate the issues associated with traditional bulk etching. As shown in Ref. \cite{hryhorenko:srf23-wepwb050}, the cavity fabrication steps do not create the damages that require bulk chemistry; rather, it is the sheet production, in particular the rolling step, that is responsible for the presence of the major crystalline damages that are expanded to the bulk. This simple modification promises not only to elevate the niobium surface quality but also to lead to greater reliability and cost-effectiveness in cavity preparation. By adding MP to the cavity production, we aim to set new standards for niobium surface treatment methodologies in the realm of superconducting radio frequency applications.

\section{\label{sec:acknowledgements}Acknowledgments}

The authors acknowledge IJCLAB Vacuum and Surfaces Platform for technical support and instrumentation resources in the surface topography evaluation, and metallographic polishing. We are also grateful to the HZB staff for assisting with the RF tests. This work was coauthored by Jefferson Science Associates LLC under U.S. DOE Contract No. DE-AC05-06OR23177. Part of this work was supported by the European Nuclear Science and Application Research-2 (ENSAR-2) under grant agreement N° 654002. This project has received funding from the European Union’s Horizon 2020 Research and Innovation programme (IFAST) under Grant Agreement No 101004730 to perform the RF tests, electropolishing, and high-temperature annealing under vacuum. 

\bibliographystyle{unsrt}  
\bibliography{main}

\end{document}